# The Importance of Tight f Basis Functions for Heavy p-Block Oxides and Halides: A Parallel With Tight d functions in the Second Row

*Published as part of The Journal of Physical Chemistry A virtual special issue "Krishnan Raghavachari Festschrift".*

Nisha Mehta* and Jan M. L. Martin*

**ABSTRACT:** It is well-known that both wave function ab initio and DFT calculations on second-row compounds exhibit anomalously slow basis set convergence unless the basis sets are augmented with additional "tight" (high-exponent) d functions, as in the cc-pV(n+d)Z and aug-cc-pV(n+d)Z basis sets. This has been rationalized as being necessary for a better description of the low-lying 3d orbital, which as the oxidation state increases sinks low enough to act as a back-donation acceptor from chalcogen and halogen lone pairs. This prompts the question whether a similar phenomenon exists for the isovalent compounds of the heavy p-block. We show that for the fourth and fifth row, this is the case, but this time for tight f functions enhancing the description of the low-lying 4f and 5f Rydberg orbitals, respectively. In the third-row heavy p block, the 4f orbitals are too far up, while the 4d orbitals are adequately covered by the basis functions already present to describe the 3d subvalence orbitals.

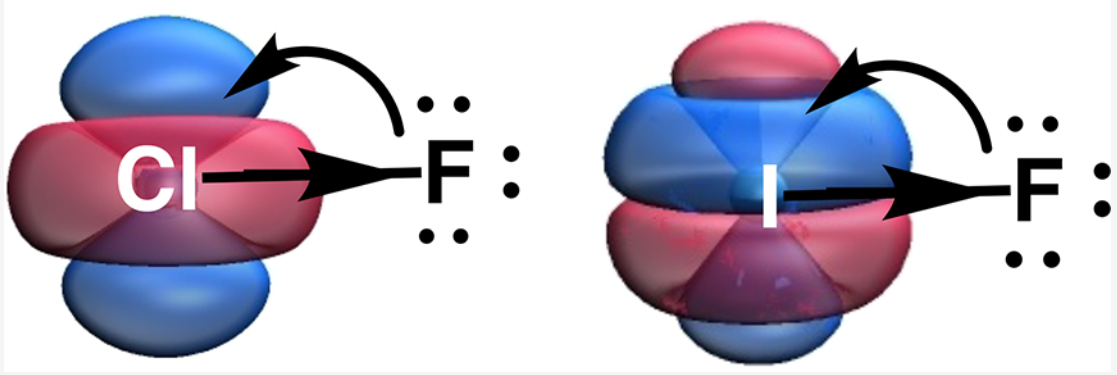

## 1. INTRODUCTION

When, in the early 1990s, the G1 and G2 computational thermochemistry approaches[1] were extended to second-row elements,[2,3] $SO_2$ was found to be a significant outlier. The G1 team found that adding a third set of d functions to the basis set increased the atomization energy of $SO_2$ by 8 kcal/mol; they ascribed this to hypervalence. Indeed in G4 theory,[4] an additional layer of d functions is placed on Al-Cl.

Bauschlicher and Partridge[5] studied basis set convergence for $SO_2$ in detail for both CCSD(T) and the B3LYP density functional approach, and at both levels found hypersensitivity to high-exponent d functions.

Martin,[6] in 1998, showed that this is the case also for other properties such as vibrational frequencies, as well as that nearly the entire contribution in CCSD(T) can be ascribed to the HF-SCF component of the energy. Moreover, this persisted when the inner-shell orbitals were replaced by an effective core potential, refuting claims that core polarization might be involved.

Later benchmark studies on second-row molecules found more severe examples in $SO_3$ (40 kcal/mol)[7] and in perchloric acid and perchloric anhydride (50 and 100 kcal/mol, respectively).[8] It was also found (see ref 8 for a discussion) that the strength of the effect was roughly proportionate to the formal oxidation state of the central second-row atom: NBO (natural bond orbital) analysis[9] revealed[8] that, as the central atom becomes more positively charged, the 3d orbitals sink lower and become ever more available for back-donation from chalcogen and halogen lone pairs.

The textbook concept of hypervalence, e.g., $d^2sp^3$ hybridization in $SF_6$ which violates the octet rule, has been comprehensively refuted by Reed and Schleyer[10] and by Cioslowski and Mixon[11] (see Norman and Pringle[12] for a recent review, as well as section 7.4 of Schwerdtfeger, Frenking, and co-workers[13]). One of us has, however, referred[14] to 3d orbitals in such molecules as "honorary valence orbitals of the second kind". (The "first kind" in that paper refers to subvalence orbitals that are energetically so close to the valence shell that freezing them in correlated calculations may cause catastrophic errors.[15,16])

In response to the findings described above, the correlation consistent basis sets for second-row elements have been revised[17] to include additional "tight" (i.e., high-exponent) d functions, giving rise to cc-pV(n+d)Z and aug-cc-pV(n+d)Z basis sets.

The authors have both been asked whether this situation is unique to second-row elements, and if yes, why. We shall show below that, in fact, a similar phenomenon occurs for 4f orbitals in the fourth-row heavy p-block elements—the last p-row of the periodic table before the lanthanides—as well as for 5f orbitals in the fifth-row heavy p-block—the last p-row before the actinides. To the best of our knowledge, the first work to refer to the need for tight f functions was the 2005 Weigend−Ahlrichs paper defining the "def2" basis sets.[18] Studies by Dixon et al.[19] on iodine fluorides and their ions, and by Peterson[20] on IOO−, both reference this same necessity, while Hill and Peterson,[21] in the paper introducing the cc-VnZ-PP-F12 basis sets for the heavy p-block elements, add 1d1f tight functions for the same reason as well as for describing outer-core correlation. However, while the need for tight f functions on heavy p-block elements has been pointed out prior to this work, it is safe to say it is not broadly known in the theoretical chemistry community and has not been investigated systematically.

**Table 1. Difference between the Hartree−Fock Component of the Total Atomization Energy Obtained with the aVTZ(-PP) and aVTZ(-PP)+(d,f) Basis Sets [aVTZ(-PP)-aVTZ(-PP)+(d,f) or awCVTZ-PP in kcal/mol][a]**

| | | Central atom | | | Central atom | | | Central atom | | | Central atom | |
|---|---|---|---|---|---|---|---|---|---|---|---|---|
| | | Rydberg | | | Rydberg | | | Rydberg | | | Rydberg | |
| | ΔSCF | f | d | ΔSCF | f | d | ΔSCF | f | d | ΔSCF | f | d |
| | | $ClO_4^-$ | | | $SO_3$ | | | $PF_5$ | | | | |
| aVTZ | REF | 0.0097 | 0.2806 | REF | 0.0107 | 0.1885 | REF | 0.0163 | 0.1226 | | | |
| aV(T+d)Z | 21.20 | 0.0095 | 0.3220 | 14.94 | 0.0106 | 0.2176 | 9.53 | 0.0161 | 0.1382 | | | |
| aVTZ+1d | 20.29 | 0.0095 | 0.3206 | 14.50 | 0.0105 | 0.2159 | 9.26 | 0.0160 | 0.1373 | | | |
| aVTZ+2d | 23.57 | 0.0095 | 0.3275 | 16.67 | 0.0106 | 0.2207 | 10.65 | 0.0161 | 0.1401 | | | |
| aVTZ+1f | 0.72 | 0.0102 | 0.2805 | 0.71 | 0.0108 | 0.1884 | 1.12 | 0.0167 | 0.1226 | | | |
| aVTZ+2d1f | 24.34 | 0.0100 | 0.3275 | 17.44 | 0.0107 | 0.2207 | 11.81 | 0.0165 | 0.1401 | | | |
| awCVTZ | 24.49 | 0.0101 | 0.3265 | 17.52 | 0.0104 | 0.2202 | 11.89 | 0.0160 | 0.1395 | | | |
| | | $BrO_4^-$ | | | $SeO_3$ | | | $AsF_5$ | | | $KrF_6$ | |
| aVTZ-PP | REF | 0.0185 | 0.1367 | REF | 0.0163 | 0.0841 | REF | 0.0187 | 0.0521 | REF | 0.0153 | 0.1379 |
| aVTZ-PP+1d | 0.67 | 0.0186 | 0.1388 | 0.49 | 0.0163 | 0.0859 | 0.34 | 0.0188 | 0.0533 | 0.29 | 0.0153 | 0.1394 |
| aVTZ-PP+2d | 0.73 | 0.0187 | 0.1395 | 0.51 | 0.0164 | 0.0866 | 0.37 | 0.0188 | 0.0537 | 0.35 | 0.0153 | 0.1396 |
| aVTZ-PP+1f | 1.54 | 0.0202 | 0.1364 | 1.19 | 0.0171 | 0.0839 | 1.65 | 0.0199 | 0.0520 | 0.38 | 0.0159 | 0.1378 |
| aVTZ-PP+2d1f | 2.30 | 0.0203 | 0.1393 | 1.73 | 0.0172 | 0.0865 | 2.05 | 0.0200 | 0.0536 | 0.74 | 0.0159 | 0.1396 |
| aVTZ-PP+2f | 1.62 | 0.0204 | 0.1364 | 1.24 | 0.0172 | 0.0839 | 1.71 | 0.0201 | 0.0520 | 0.43 | 0.0159 | 0.1378 |
| aVTZ-PP+2d2f | 2.38 | 0.0205 | 0.1393 | 1.78 | 0.0173 | 0.0865 | 2.11 | 0.0201 | 0.0535 | 0.78 | 0.0159 | 0.1396 |
| awCVTZ-PP | 2.80 | 0.0203 | 0.1384 | 2.16 | 0.0171 | 0.0859 | 2.97 | 0.0197 | 0.0530 | 0.98 | 0.0161 | 0.1398 |
| | | $IO_4^-$ | | | $TeO_3$ | | | $SbF_5$ | | | $XeF_6$ | |
| aVTZ-PP | REF | 0.0416 | 0.0842 | REF | 0.0261 | 0.0529 | REF | 0.0222 | 0.0280 | REF | 0.0343 | 0.1022 |
| aVTZ-PP+1d | 0.08 | 0.0419 | 0.0867 | 0.08 | 0.0262 | 0.0545 | 0.10 | 0.0222 | 0.0288 | 0.03 | 0.0343 | 0.1046 |
| aVTZ-PP+2d | 0.16 | 0.0420 | 0.0879 | 0.09 | 0.0263 | 0.0552 | 0.12 | 0.0223 | 0.0290 | 0.13 | 0.0343 | 0.1059 |
| aVTZ-PP+1f | 7.98 | 0.0554 | 0.0829 | 4.47 | 0.0325 | 0.0524 | 5.31 | 0.0288 | 0.0276 | 3.95 | 0.0426 | 0.1015 |
| aVTZ-PP+2d1f | 8.15 | 0.0558 | 0.0866 | 4.58 | 0.0326 | 0.0547 | 5.42 | 0.0288 | 0.0284 | 4.07 | 0.0426 | 0.1052 |
| aVTZ-PP+2f | 11.36 | 0.0660 | 0.0824 | 5.88 | 0.0367 | 0.0523 | 6.71 | 0.0320 | 0.0275 | 6.63 | 0.0503 | 0.1009 |
| aVTZ-PP+2d2f | 11.52 | 0.0663 | 0.0861 | 5.98 | 0.0368 | 0.0546 | 6.83 | 0.0320 | 0.0285 | 6.75 | 0.0503 | 0.1047 |
| awCVTZ-PP | 12.72 | 0.0660 | 0.0852 | 7.28 | 0.0367 | 0.0542 | 9.55 | 0.0318 | 0.0280 | 8.09 | 0.0507 | 0.1049 |
| | | $AtO_4^-$ | | | $PoO_3$ | | | | | | | |
| aVTZ-PP | REF | 0.0439 | 0.0434 | REF | 0.0258 | 0.0261 | | | | | | |
| aVTZ-PP+1d | 0.07 | 0.0442 | 0.0456 | 0.02 | 0.0259 | 0.0275 | | | | | | |
| aVTZ-PP+2d | 0.20 | 0.0442 | 0.0462 | 0.08 | 0.0259 | 0.0279 | | | | | | |
| aVTZ-PP+1f | 8.38 | 0.0608 | 0.0425 | 4.76 | 0.0336 | 0.0258 | | | | | | |
| aVTZ-PP+2d1f | 8.55 | 0.0610 | 0.0453 | 4.84 | 0.0337 | 0.0275 | | | | | | |
| aVTZ-PP+2f | 9.98 | 0.0673 | 0.0424 | 5.49 | 0.0366 | 0.0257 | | | | | | |
| aVTZ-PP+2d2f | 10.15 | 0.0675 | 0.0452 | 5.57 | 0.0367 | 0.0274 | | | | | | |
| awCVTZ-PP | 11.73 | 0.0675 | 0.0447 | 7.16 | 0.0367 | 0.0270 | | | | | | |

[a]The NPA population of the d and f Rydberg orbitals at the Hartree−Fock level for the given basis set is also reported.

## 2. METHODS

In the present work, we investigate the effect of tight d and f functions on the total atomization energies of heavy p-block oxides and halides at the HF, DFT, and CCSD(T)[22,23] level of theories. All geometries are optimized at the PW6B95[24]-D3(BJ)[25]/def2-QZVPP[18] level of theory, and XYZ coordinates are reported in the Supporting Information. All calculations are carried out using the Gaussian 16[26] and MOLPRO[27] Program suites, running on the Faculty of Chemistry HPC facility "ChemFarm" at the Weizmann Institute. We considered the correlation consistent basis sets of Dunning and co-workers.[28,29] In this paper, we considered aug-cc-pVnZ[30] and aug-cc-pV(n+d)Z[17] basis sets for second-row elements and aug-cc-pVnZ-PP[31,32] for heavy p-block elements, where PP stands for the Stuttgart−Cologne energy-consistent relativistic pseudopotentials.[33] We also considered weighted core−valence basis sets: aug-cc-pwCVnZ[30,34] for second-row and aug-cc-pwCVnZ-PP[31,32,35] for heavy p-block elements. For brevity, aug-cc-pVnZ(-PP), aug-cc-pV(n+d)Z, and aug-cc-pwCVnZ(-PP) basis sets are denoted as aVnZ(-PP), aV(n+d)Z, and awCVnZ(-PP), respectively. The high-exponent (d,f) functions that are added to the aVnZ(-PP) basis set are taken from the core−valence awCVnZ(-PP) basis sets for the respective elements. All calculations are carried out within the frozen core approximation with Gaussian's `int(grid = ultrafine)`, which corresponds to the pruned direct product of a 99-point Euler-Maclaurin radial and a 590-point Lebedev angular grid or its MOLPRO counterpart. The default `tight` SCF convergence criterion was used, which corresponds to the norm of the density matrix update being $<10^{-8}$ and the largest individual element change being $<10^{-6}$ in absolute value. We also carried out natural bond orbital (NBO) population analysis[9] at the Hartree−Fock and DFT levels using the NBO7 program[36] interfaced to both Gaussian 16[26] and MOLPRO.[27] As Gaussian has no CCSD(T) analytical derivatives and hence cannot write out first-order reduced density matrices at the CCSD(T) level, we assessed the importance of (T) for the NBO populations using MOLPRO, both using its built-in NBO implementation and using NBO7 via the interface. The differences between CCSD and CCSD(T) NBO populations were found to be negligible. The BFGS algorithm[37] as implemented in MOLPRO 2022 is used for the exponent optimizations, with the gradient threshold set to $10^{-5}$.

## 3. RESULTS AND DISCUSSION

### 3.1. Effect of Tight d and f Functions on the Hartree−Fock Component.

Table 1 presents the difference between the atomization energies calculated with the aVTZ(-PP) basis set and those obtained by adding a progressively large set of (d,f) functions to the basis set at the Hartree−Fock level. As it was previously found (e.g., Section 2.2 in ref 38) for the second-row elements that core−valence basis functions are already in the right exponent range, we assumed that the tight d and f exponents from the cc-pwCVnZ basis sets were a reasonable starting point.

First, we analyze the $XO_4^-$ series (where X = Cl, Br, I, and At). The effect of adding tight d functions to the aVTZ basis set is quite significant for $ClO_4^-$, as well-documented:[8] for instance, adding one d function to the aVTZ basis set increases the atomization energy of $ClO_4^-$ (i.e., $\Delta TAE_{SCF}$) by 20.29 kcal/mol. The corresponding values for adding progressively larger sets of tight (d,f) functions to the aVTZ basis set are 23.57 (+2d), 0.72 (+1f), and 24.34 (+2d1f) kcal/mol. Therefore, the large TAE increase seen for aug-cc-pwCVTZ (24.49 kcal/mol) is almost entirely due to the effect of d functions on the Cl atom. Table 1 also lists the overall population of the d and f Rydberg orbitals taken from the NBO population analysis of the HF determinant. (This can be obtained either by modifying the NBO7 source code to print more than two decimal places in the `Natural Electron Configuration` section of the output or by means of a simple shell and `awk` script that searches for the Rydberg NAO occupations of a given angular momentum and sums them up.) An NBO analysis of the wave function reveals that the natural population of chlorine d orbitals ($q_{3d}$) increases as much as 0.05 when tight d functions are added. Specifically, $\delta q_{3d}$ values, that is, the changes in the NPA populations relative to the aVTZ basis set, are 0.041, 0.040, and 0.047, respectively, for aV(T+d)Z, aVTZ+1d, and aVTZ+2d. As discussed in the Introduction, the chemical significance of the tight d functions to the aVTZ basis set is that they increase the ability of the chlorine 3d orbitals (in $ClO_4^-$) to act as back-bonding acceptors. The chlorine 4f orbitals are too far up in energy to participate significantly. These findings are consistent with those reported in ref 8 for $Cl_2O_7$ and $HClO_4$.

The effect of high-exponent d functions is drastically reduced for third-row pseudohypervalent molecules (e.g., $BrO_4^-$). Adding two high-exponent d functions to the aVTZ-PP basis set affects the SCF component of the total atomization energy by a measly 0.73 kcal/mol, compared to 23.57 kcal/mol for $ClO_4^-$. Likewise, $q_{4d}$ is affected by only 0.003. Furthermore, inclusion of tight f functions (i.e., aVTZ-PP+2f) has a negligible effect, and the HF-SCF component of the total atomization energy is affected by just 1.62 kcal/mol, which is considerably less than other sources of basis set incompleteness error for aVTZ-PP. Furthermore, the $4f_{pop}$ population is essentially unchanged. In the case of $BrO_4^-$, the bromine 3d orbitals are filled. Bromine 4d orbitals can act as back-bonding acceptors, but enough high-exponent d functions are already present in the underlying basis set (for the purpose of describing the 3d subvalence orbital); hence, the additional high-exponent d functions have a negligible contribution. The 4f orbitals are still too far up; hence, they do not benefit from the additional tight f functions.

Next, what happens when high-exponent (d,f) functions are added to the aVTZ-PP basis set for the $IO_4^-$ molecule? Analogous to perbromate, adding tight d functions for $IO_4^-$ affects the HF-SCF(TAE) value by less than 0.2 kcal/mol, making them chemically insignificant. The situation for high-exponent f functions is starkly different: $TAE_{SCF}$ values are increased by 7.98 (aVTZ-PP+1f) and 8.15 (aVTZ-PP+2f) kcal/mol, with a concomitant $q_{4f}$ increase of ≈0.02. While the iodine 4d orbitals are filled, 5d and 4f orbitals have nontrivial occupations. Similarly to $BrO_4^-$, there are enough decontracted functions from the valence shell of the basis set for the iodine 5d orbitals, but extra primitives are indeed necessary for 4f orbitals. Analogous to 3d orbitals in $ClO_4^-$, the addition of extra high-exponent f functions on iodine increases the ability of low-lying virtual 4f orbitals to act as back-bonding acceptors from chalcogen and halogen lone pairs.

The effect of high-exponent f functions is even more pronounced in perastatate, $AtO_4^-$. For instance, TAE[$AtO_4^-$] is increased by appreciable amounts of 8.38 (aVTZ-PP+1f) and 9.98 (aVTZ+2f) kcal/mol. The extra f functions increase $q_{5f}$ on the At atom by ≈0.02. As expected, only trivial increases are seen for aVTZ+1d (0.07 kcal/mol) and aVTZ+2d (0.20 kcal/mol) basis sets. While the valence 5d orbitals are filled, 6d and 5f orbitals have significant occupations. Again, while there are enough decontracted functions for 6d orbitals from the valence shell of the basis set, extra high-exponent f functions are needed for the description of 5f orbitals.

We can summarize the above findings by saying that, for p-block elements in higher oxidation states, the second row requires tight d functions, while the fourth and fifth rows require tight f functions, and the third row requires neither. The fact that the second-row heavy p-block is approaching the first-row transition elements, and the fourth and fifth row heavy p-blocks the lanthanide and actinide series, respectively, is not a coincidence but directly linked to the energetic proximity of the d and f Rydberg orbitals.

For systems having lower oxidation states on the central atom, proportionally smaller effects are observed. For $ClO_3^-$ and $ClO_2^-$, the HF-SCF component of the total atomization energy is affected by 12.909 and 5.113 kcal/mol, respectively, whereas for $IO_3^-$, $AtO_3^-$, $IO_2^-$, and $AtO_2^-$, the effect of the tight f functions ranges from 2.665 to 7.250 kcal/mol (Table 2).

What happens to the equilibrium distances and harmonic frequencies when extra (d,f) functions are added? Table 3 contains the computed $r$(X−O) bond distances and vibrational frequencies for $XO_4^-$ (where X = Cl, Br, I, and At) obtained by progressively adding (d,f) functions to the aVTZ basis set. There are only four unique vibrational frequencies: $T_2$ (triply degenerate bend), E (degenerate bend), $A_1$ (symmetric stretch), and $T_2$ (triply degenerate asymmetric stretch). Adding just one d exponent shortens the $r$(Cl−O) distance by 0.013 Å, and consequently,[39,40] the vibrational frequencies are blue-shifted by 3−4% (16.6, 23.3, 42.9, and 31.0 cm$^{-1}$, respectively). aVTZ+2d further raises the vibrational frequencies by 2.5, 3.6, 6.9, and 5.2 cm$^{-1}$. The addition of high-exponent f functions, on the other hand, has an insignificant impact on $r$(Cl−O) distance and vibrational frequencies, as expected.

**Table 2. Effect of Tight d and f Functions on the HF-SCF Component of the TAE (kcal/mol) as a Function of the Oxidation State of the Central Halogen Atom**

| | aVnZ vs aV(n+d)Z, aVnZ-PP vs aVnZ-PP+1f | | | | | |
|---|---|---|---|---|---|---|
| | n = D | | n = T | | n = Q | |
| | d | f | d | f | d | f |
| $ClO_4^-$ | 40.408 | | 21.205 | | 12.091 | |
| $ClO_3^-$ | 24.087 | | 12.909 | | 7.448 | |
| $ClO_2^-$ | 9.420 | | 5.113 | | 3.013 | |
| $ClO^-$ | 2.140 | | 1.156 | | 0.698 | |
| $BrO_4^-$ | 7.178 | 1.730 | 0.672 | 1.542 | 0.105 | 0.500 |
| $BrO_3^-$ | 4.646 | 1.379 | 0.405 | 1.174 | 0.082 | 0.424 |
| $BrO_2^-$ | 1.825 | 0.599 | 0.151 | 0.502 | 0.042 | 0.200 |
| $BrO^-$ | 0.460 | 0.185 | 0.033 | 0.148 | 0.014 | 0.067 |
| $IO_4^-$ | | 8.882 | | 7.984 | | 4.073 |
| $IO_3^-$ | | 6.763 | | 6.136 | | 3.064 |
| $IO_2^-$ | | 2.892 | | 2.665 | | 1.343 |
| $IO^-$ | | 0.908 | | 0.829 | | 0.419 |
| $AtO_4^-$ | | 14.110 | | 8.384 | | 4.295 |
| $AtO_3^-$ | | 11.650 | | 7.250 | | 3.599 |
| $AtO_2^-$ | | 4.936 | | 3.105 | | 1.545 |
| $AtO^-$ | | 1.566 | | 0.972 | | 0.482 |

Proceeding to $BrO_4^-$, the addition of inner polarization d and/or f functions has a modest effect on $r$(Br−O) bond distances and vibrational frequencies (just 2−3 cm$^{-1}$).

In contrast, for the iodine and astatine oxides, adding a high-exponent f function blue-shifts all frequencies, by 5.8, 3.5, 12, and 11.4 cm$^{-1}$ for $IO_4^-$ and by 5.8, 2.7, 14.4, and 12.7 cm$^{-1}$ for $AtO_4^-$. Adding a second hard f function still contributes 1.8, 1.1, 5, and 4.9 cm$^{-1}$ for $IO_4^-$, which is somewhat significant for the stretching frequencies, and 0.9, 0.3, 2.5, and 2.3 cm$^{-1}$ for $AtO_4^-$. The tight d exponent contributions for $IO_4^-$ and $AtO_4^-$ molecules are essentially nil.

As expected, and consistent with the lower oxidation states of the central atoms, for the chalcogen and pnictogen based molecules $SO_3$, $SeO_3$, $TeO_3$, $PoO_3$, $PF_5$, $AsF_5$, and $SbF_5$, we obtain somewhat milder effects of the tight (d,f) functions. Consistent with the halogen series, the addition of two high-exponent d functions increases the atomization energy of $SO_3$ by 16.67 kcal/mol; for $PF_5$, the magnitude of the effect is 10.65 kcal/mol. The contribution of the hard d exponent dwindles from right to left in the third row of the Periodic Table: from 0.49 ($SeO_3$) to 0.34 kcal/mol ($AsF_5$). Note also that adding a second hard d function contributes negligibly (0.03 kcal/mol). For $TeO_3$, $PoO_3$, and $SbF_5$, as advocated in the previous section, the high-exponent f functions' contribution increases rapidly for the third and fourth-row pseudohypervalent molecules. The addition of one high-exponent f function increases the total

**Table 3. Effect of Tight (d,f) Functions on Equilibrium Bond Distances (Å) and Harmonic Frequencies (cm$^{-1}$) at the HF Level**

$ClO_4^-$

| | aVTZ | aVTZ+1d | aVTZ+2d | aVTZ+1f | aVTZ+2d1f | awCVTZ | aV(T+d)Z |
|---|---|---|---|---|---|---|---|
| $\omega_1$(E) | 502.3 | 518.9 | 521.4 | 503.4 | 522.7 | 522.5 | 519.3 |
| $\omega_2$($T_2$) | 696.0 | 719.3 | 722.9 | 697.2 | 724.3 | 724.2 | 720.1 |
| $\omega_3$($A_1$) | 1034.6 | 1077.5 | 1084.4 | 1035.7 | 1085.4 | 1085.3 | 1080.4 |
| $\omega_4$($T_2$) | 1212.5 | 1243.5 | 1248.7 | 1212.6 | 1248.5 | 1247.9 | 1246.9 |
| $r$(Cl−O) | 1.428 | 1.415 | 1.412 | 1.427 | 1.412 | 1.412 | 1.414 |

$BrO_4^-$

| | aVTZ-PP | aVTZ-PP+1d | aVTZ-PP+2d | aVTZ-PP+1f | aVTZ-PP+2d1f | aVTZ-PP+2f | aVTZ-PP+2d2f | awCVTZ-PP |
|---|---|---|---|---|---|---|---|---|
| $\omega_1$(E) | 383.1 | 383.7 | 383.7 | 385.1 | 385.8 | 385.1 | 385.9 | 386.1 |
| $\omega_1$($T_2$) | 486.5 | 487.4 | 487.5 | 488.2 | 489.2 | 488.2 | 489.2 | 489.5 |
| $\omega_1$($A_1$) | 920.0 | 921.6 | 921.7 | 922.9 | 924.6 | 923.1 | 924.8 | 925.6 |
| $\omega_1$($T_2$) | 1012.7 | 1013.5 | 1013.7 | 1014.9 | 1015.8 | 1015.1 | 1015.9 | 1016.3 |
| $r$(Br−O) | 1.578 | 1.577 | 1.577 | 1.575 | 1.575 | 1.575 | 1.574 | 1.574 |

$IO_4^-$

| | aVTZ-PP | aVTZ-PP+1d | aVTZ-PP+2d | aVTZ-PP+1f | aVTZ-PP+2d1f | aVTZ-PP+2f | aVTZ-PP+2d2f | awCVTZ-PP |
|---|---|---|---|---|---|---|---|---|
| $\omega_1$(E) | 301.9 | 302.0 | 302.1 | 307.7 | 307.8 | 309.5 | 309.6 | 309.8 |
| $\omega_2$($T_2$) | 368.0 | 368.1 | 368.2 | 371.5 | 371.8 | 372.6 | 372.9 | 373.0 |
| $\omega_3$($A_1$) | 890.5 | 890.5 | 890.7 | 902.5 | 902.5 | 907.5 | 907.5 | 909.4 |
| $\omega_4$($T_2$) | 956.2 | 956.0 | 956.2 | 967.6 | 967.4 | 972.5 | 972.3 | 972.9 |
| $r$(I−O) | 1.747 | 1.747 | 1.747 | 1.737 | 1.736 | 1.733 | 1.732 | 1.731 |

$AtO_4^-$

| | aVTZ-PP | aVTZ-PP+1d | aVTZ-PP+2d | aVTZ-PP+1f | aVTZ-PP+2d1f | aVTZ-PP+2f | aVTZ-PP+2d2f | awCVTZ-PP |
|---|---|---|---|---|---|---|---|---|
| $\omega_1$(E) | 251.0 | 251.0 | 251.1 | 256.8 | 256.9 | 257.7 | 257.7 | 257.9 |
| $\omega_2$($T_2$) | 281.6 | 281.6 | 281.7 | 284.3 | 284.4 | 284.6 | 284.7 | 285.1 |
| $\omega_3$($A_1$) | 760.5 | 760.5 | 760.7 | 774.9 | 774.9 | 777.4 | 777.4 | 780.8 |
| $\omega_4$($T_2$) | 800.9 | 800.7 | 801.0 | 813.6 | 813.4 | 815.9 | 815.7 | 817.8 |
| $r$(At−O) | 1.866 | 1.866 | 1.866 | 1.851 | 1.851 | 1.848 | 1.848 | 1.846 |

**Table 4. Comparison of Optimized Tight f Exponents with Those Taken from the Core−Valence Basis Sets of Iodine and Astatine and Comparison of ΔTAE**

| | | | ΔTAE (kcal/mol) | | | ΔTAE (kcal/mol) | |
|---|---|---|---|---|---|---|---|
| | | exponent | HF | CCSD(T) | exponent | HF | CCSD(T) |
| | | from awCVTZ or awCVTZ-PP | | | SCF optimization in $XO_4^-$ | | |
| $IO_4^-$ | f | 1.393 | 7.984 | 8.368 | 1.692 | 8.215 | 8.735 |
| | 2f | 1.393, 4.867 | 11.360 | 12.365 | 1.329, 5.557 | 11.582 | 12.564 |
| $AtO_4^-$ | f | 1.036000 | 8.384 | 9.979 | 1.340 | 8.975 | 10.927 |
| | 2f | 1.036, 2.704 | 9.975 | 12.270 | 0.853, 2.450 | 10.071 | 12.404 |
| | | from awCVQZ or awCVQZ-PP | | | SCF optimization in $XO_4^-$ | | |
| $IO_4^-$ | f | 1.470 | 4.073 | 4.672 | 2.779 | 5.902 | 6.786 |
| | 2f | 1.470, 4.102 | 7.240 | 8.355 | 2.200, 8.451 | 8.162 | 9.441 |
| $AtO_4^-$ | f | 1.193 | 4.294 | 5.769 | 1.846 | 5.396 | 7.228 |
| | 2f | 1.193, 2.032 | 5.499 | 7.373 | 1.334, 3.085 | 5.630 | 7.559 |
| | | from awCV5Z or awCV5Z-PP | | | SCF optimization in $XO_4^-$ | | |
| $IO_4^-$ | f | 1.500 | 3.132 | 3.650 | 3.264 | 5.317 | 6.176 |
| | 2f | 1.500, 3.439 | 5.722 | 6.657 | 2.560, 9.633 | 7.261 | 8.477 |
| $AtO_4^-$ | f | 1.313 | 3.224 | 4.409 | 2.067 | 4.043 | 5.498 |
| | 2f | 1.313, 2.198 | 4.084 | 5.561 | 1.529, 3.372 | 4.156 | 5.665 |

**Table 5. Changes in NBO Occupations on the Central Halogen Atom upon the Addition of an Extra d or f Function to the aVTZ(-PP) Basis Set and Increases in the Total Atomization Energy (TAE, kcal/mol) upon Addition of Said Functions**

| | Δ*q*(d) | | | Δ*q*(f) | | |
|---|---|---|---|---|---|---|
| | HF-SCF | $CCSD(T)_{corr}$ | (T) | HF-SCF | $CCSD(T)_{corr}$ | (T) |
| $ClO_4^-$ | 0.0405 | 0.0084 | 0.0016 | 0.0004 | 0.0003 | 0.0001 |
| $BrO_4^-$ | 0.0021 | 0.0005 | 0.0001 | 0.0017 | 0.0004 | 0.0001 |
| $IO_4^-$ | 0.0025 | 0.0000 | 0.0000 | 0.0138 | 0.0021 | 0.0002 |
| $AtO_4^-$ | 0.0003 | −0.0004 | −0.0001 | 0.0169 | 0.0047 | 0.0007 |
| | | | ΔTAE (kcal/mol) | | | |
| | | | aVTZ vs aV(T+d)Z, aVTZ-PP vs aVTZ-PP+1f | | | |
| $ClO_4^-$ | 21.205 | 1.924 | 0.318 | 0.717 | 0.110 | 0.008 |
| $BrO_4^-$ | 0.672 | 0.186 | 0.015 | 1.542 | −0.120 | 0.003 |
| $IO_4^-$ | 0.083 | 0.023 | 0.004 | 7.984 | 0.377 | 0.007 |
| $AtO_4^-$ | 0.071 | −0.014 | 0.011 | 8.384 | 1.595 | 0.181 |
| | | | aVQZ vs aV(Q+d)Z, aVQZ-PP vs aVQZ-PP+1f | | | |
| $ClO_4^-$ | 12.091 | 1.175 | 0.204 | | | |
| $IO_4^-$ | | | | 4.073 | 0.600 | 0.001 |
| $AtO_4^-$ | | | | 4.294 | 1.475 | 0.120 |
| | | | aV5Z vs aV(5+d)Z, aV5Z-PP vs aV5Z-PP+1f | | | |
| $ClO_4^-$ | 2.697 | 0.184 | 0.042 | | | |
| $IO_4^-$ | | | | 3.650 | 0.518 | 0.008 |
| $AtO_4^-$ | | | | 3.224 | 1.185 | 0.112 |
| | | | aV6Z vs aV(6+d)Z | | | |
| $ClO_4^-$ | 0.946 | 0.039 | 0.017 | | | |
| | | | aVDZ vs aV(D+d)Z, aVDZ-PP vs aVDZ-PP+1f | | | |
| $ClO_4^-$ | 40.408 | 2.380 | 0.524 | | | |
| $BrO_4^-$ | 7.178 | 0.266 | 0.110 | 1.730 | −0.008 | 0.009 |
| $IO4_4^-$ | | | | 8.882 | 0.359 | 0.010 |
| $AtO_4^-$ | | | | 14.110 | 3.232 | 0.254 |

atomization energy of $TeO_3$ and $SbF_5$ by around 5 kcal/mol; for two f functions, the magnitude of effect is further increased by 1−2 kcal/mol.

A remark is in order about the noble gas fluorides (e.g., $KrF_6$ and $XeF_6$). Consistent with $BrO_4^-$, little effect is seen (see Table 1) from addition of either tight d or f functions to the Kr aVTZ-PP basis set. For $XeF_6$ in aVTZ-PP, on the other hand, the first added tight f function accounts for 3.95 kcal/mol at the HF level. (As expected, we find that additional tight d functions have no impact worth speaking of.) Adding the next f function increases $TAE_{SCF}$ by 2.68 kcal/mol: clearly, this effect is not specific to halogens.

In order to verify that our conclusions are not due to an artifact of our choice of exponents (i.e., tight d and f functions taken from the core−valence basis sets for the respective elements), we also reoptimized them at the HF level in the respective perhalides $ClO_4^-$, $BrO_4^-$, $IO_4^-$, and $AtO_4^-$. The exponents we found are reported in Table 4. Our main point remains unchanged; however, for higher angular momentum basis sets, such as aVQZ-PP and aV5Z-PP, a single optimized f exponent recovers nearly as much energy as two fixed ones from the awCVQZ(-PP) and awCV5Z-PP basis sets, respectively, especially for $AtO_4^-$. For aVTZ-PP basis set, the exponents we found are actually, by and large, close to those obtained from the awCVTZ-PP basis set.

Table 6. Difference between the $E_{corr}$CCSD and $E_{corr}$(T) Components of the Total Atomization Energy Obtained with the aVTZ(-PP) and aVTZ(-PP)+(d,f) Basis Sets [aVTZ(-PP)-aVTZ(-PP)+(d,f) or awCVTZ-PP in kcal/mol][a]

| | $\Delta E_{corr}$CCSD | $\Delta E_{corr}$(T) | $\Delta E_{corr}$CCSD | $\Delta E_{corr}$(T) | $\Delta E_{corr}$CCSD | $\Delta E_{corr}$(T) | $\Delta E_{corr}$CCSD | $\Delta E_{corr}$(T) |
|---|---|---|---|---|---|---|---|---|
| | $ClO_4^-$ | | $SO_3$ | | $PF_5$ | | | |
| aVTZ | REF | REF | REF | REF | REF | REF | | |
| aV(T+d)Z | 1.606 | 0.317 | 0.359 | 0.005 | 0.403 | 0.057 | | |
| aVTZ+1d | 1.704 | 0.321 | 0.479 | 0.016 | 0.486 | 0.062 | | |
| aVTZ+2d | 1.915 | 0.380 | 0.529 | 0.023 | 0.551 | 0.073 | | |
| aVTZ+1f | 0.102 | 0.008 | 0.220 | 0.021 | 0.175 | 0.005 | | |
| aVTZ+2d1f | 2.168 | 0.401 | 0.861 | 0.052 | 0.787 | 0.082 | | |
| awCVTZ | 2.013 | 0.395 | 0.916 | 0.071 | 0.895 | 0.093 | | |
| | $BrO_4^-$ | | $SeO_3$ | | $AsF_5$ | | $KrF_6$ | |
| aVTZ-PP | REF | REF | REF | REF | REF | REF | REF | REF |
| aVTZ-PP+1d | 0.172 | 0.015 | 0.135 | 0.015 | 0.184 | 0.014 | 0.026 | 0.013 |
| aVTZ-PP+2d | 0.183 | 0.019 | 0.156 | 0.018 | 0.224 | 0.017 | 0.015 | 0.018 |
| aVTZ-PP+1f | −0.123 | 0.003 | −0.071 | 0.008 | 0.064 | 0.005 | −0.172 | 0.029 |
| aVTZ-PP+2d1f | 0.034 | 0.022 | 0.070 | 0.026 | 0.269 | 0.022 | −0.180 | 0.049 |
| aVTZ-PP+2f | −0.080 | 0.002 | −0.017 | 0.006 | 0.152 | 0.007 | −0.174 | 0.027 |
| aVTZ-PP+2d2f | 0.079 | 0.020 | 0.129 | 0.025 | 0.363 | 0.025 | −0.186 | 0.047 |
| awCVTZ-PP | 0.139 | 0.057 | 0.251 | 0.061 | 0.544 | 0.045 | −0.589 | 0.106 |
| | $IO_4^-$ | | $TeO_3$ | | $SbF_5$ | | $XeF_6$ | |
| aVTZ-PP | REF | REF | REF | REF | REF | REF | REF | REF |
| aVTZ-PP+1d | 0.024 | 0.004 | 0.068 | 0.013 | 0.323 | 0.018 | −0.031 | 0.005 |
| aVTZ-PP+2d | 0.053 | 0.008 | 0.107 | 0.017 | 0.399 | 0.023 | −0.019 | 0.009 |
| aVTZ-PP+1f | 0.377 | 0.007 | 0.010 | −0.031 | 1.261 | 0.100 | 0.294 | 0.070 |
| aVTZ-PP+2d1f | 0.413 | 0.017 | 0.107 | −0.012 | 1.588 | 0.122 | 0.253 | 0.082 |
| aVTZ-PP+2f | 0.945 | 0.060 | 0.289 | −0.031 | 2.559 | 0.193 | 0.689 | 0.141 |
| aVTZ-PP+2d2f | 0.989 | 0.071 | 0.404 | −0.010 | 3.003 | 0.227 | 0.633 | 0.152 |
| awCVTZ-PP | 0.757 | 0.150 | 0.322 | 0.035 | 3.376 | 0.276 | 0.006 | 0.210 |
| | $AtO_4^-$ | | $PoO_3$ | | | | | |
| aVTZ-PP | REF | REF | REF | REF | | | | |
| aVTZ-PP+1d | −0.025 | 0.011 | 0.061 | 0.019 | | | | |
| aVTZ-PP+2d | −0.016 | 0.013 | 0.068 | 0.021 | | | | |
| aVTZ-PP+1f | 1.413 | 0.181 | 0.882 | 0.094 | | | | |
| aVTZ-PP+2d1f | 1.381 | 0.199 | 0.947 | 0.118 | | | | |
| aVTZ-PP+2f | 2.038 | 0.257 | 1.406 | 0.132 | | | | |
| aVTZ-PP+2d2f | 2.008 | 0.276 | 1.495 | 0.161 | | | | |
| awCVTZ-PP | 1.255 | 0.372 | 1.180 | 0.211 | | | | |

[a]Connected doubles, $E$(CCSD) − $E$(SCF) ≡ $E_{corr}$CCSD; Noniterative connected triples, $E$(CCSD(T)) − $E$(CCSD) ≡ $E_{corr}$(T).

**3.2. Correlation Components.** It has been known since the late 1990s (e.g., ref 6) that the lion's share of the "tight d function effect" in second row elements is recovered at the Hartree−Fock level and that the correlation energy is only weakly affected.

For the $XO_4^-$ series, Table 5 sheds some light on the matter. We present there not only the contributions to the total atomization energy (in kcal/mol) but also the cumulative d- and f-type NAO (natural atomic orbital) populations on the central halogen atom. These latter populations were obtained using the built-in implementation of NBO (natural bond orbital) theory[9] in MOLPRO 2022.[27] In Table 6, a breakdown of the correlation contribution into CCSD and (T) can be found.

The largest correlation contribution is for the d orbital in $ClO_4^-$, 1.9 kcal/mol, and even that is an order of magnitude less than the HF component. Out of that, 0.3 kcal/mol is accounted for by the triples contribution. These correlation contributions are nearly cut in half for the next basis set in the series, aug-cc-pVQZ vs aug-cc-pV(Q+d)Z, and again the SCF contribution exceeds that of correlation by an order of magnitude.

While in computational thermochemistry protocols like W4 theory,[41] which covers first- and second-row molecules, we have always striven to include tight d functions also for the CCSD(T) correlation steps, the post-CCSD(T) steps always omit them (in part because of the steep computational cost scaling of CCSDT, CCSDT(Q), and CCSDTQ). A very recent study by Karton[42] reconsiders this aspect and indicates that the inclusion of tight d functions even in these steps may have enough of an effect to be significant in high-accuracy computational thermochemistry (see Karton[43] for a very recent review).

For periodate and perastatate, the tight f contribution is in the 8 kcal/mol range, albeit especially for perastatate with a more noticeable correlation contribution. Still, the HF contribution dominates by far.

As a rule of thumb, going up from aVTZ to aVQZ appears to cut the tight d or tight f contribution in half. Going one step further up the hierarchy to aV(5+d)Z and aV(6+d)Z, the tight d contribution for $ClO_4^-$ definitely tapers off smoothly, but contributions in excess of 3−4 kcal/mol are still seen with aV5Z-PP basis sets for $IO_4^-$ and $AtO_4^-$.

Comparison of the NBO charges with and without the extra functions reveals that, at the HF level, the aV(T+d)Z basis set for $ClO_4^-$ has about 0.04 electrons more d occupation than without the extra d, while the extra tight f functions on iodine and astatine cause increases of 0.014 and 0.017 for $IO_4^-$ and $AtO_4^-$, respectively. The contribution of correlation is a factor of 3−7 smaller, and out of that, (T) accounts for a negligible fraction. The latter means that CCSD will be generally enough for differential NBO analysis of this type—which is quite convenient, as CCSD densities are both much more economical than CCSD(T) and available in more electronic structure programs (notably, Gaussian).

For DFT calculations on heavier elements, the popular Weigend−Ahlrichs "def2" basis sets[18] are widely used owing to their availability for all elements through radon. Weigend and Ahlrichs[18] mention in passing that two additional tight f functions are needed for heavier p-block elements due to what they deem to be "core polarization". Now if this interpretation were correct, then the need for the tight f functions would go away if the subvalence electrons were all replaced by a large-core ECP. We thus repeated the $IO_4^-$ calculations using Martin and Sundermann's large-core SDB-aug-cc-pVTZ and QZ basis sets,[44] which for practical applications were superseded by the small-core "official" (aug-)cc-pVnZ-PP basis sets.[31,45] Despite iodine now having all 46 core electrons, $[Kr](4d)^{10}$, "rolled into" the ECP, it turns out we need the tight f functions just as much, which rules out core polarization. Repeating a numerical experiment that was "left on the cutting room floor" of the final published version of the 1998 $SO_2$ paper,[6] we replaced the sulfur core electrons in $SO_2$ and $SO_3$ with an ECP10MWB pseudopotential (while decontracting the s and p functions of the sulfur aug-cc-pVnZ basis set to avoid contraction mismatch with the ECP) and found that the need for tight d functions was likewise essentially the same as for the all-electron calculation. (See also the very recent large-core ccECP-cc-pV(n+d)Z basis sets of Hill and co-workers.[46])

**3.3. Density Functional Theory.** It stands to reason that the same observations made above for Hartree−Fock would also apply to other independent-particle models, specifically to DFT. For the tight d functions in the second row, this was indeed first spotted[5] at the B3LYP and CCSD(T) levels and later confirmed[6] at the HF level.

For the sake of completeness, we have repeated our analysis for the PW6B95 hybrid meta-GGA functional. As can be seen in the Supporting Information, our observations at the PW6B95 level are fundamentally the same as at the HF level.

## 4. CONCLUSIONS

We have examined the effect of tight d and f functions in aug-cc-pVnZ (or aug-cc-pV(n+d)Z) basis sets on SCF and post HF contributions to the total atomization energies and vibrational frequencies for p-block fluorides and oxides. From the present study, we can conclude that the need for added high-exponent d functions to the second-row p-block elements (as done in the (aug-)cc-pV(n+d)Z basis sets[17]) has a direct (albeit milder) parallel in the fourth and fifth row, but now in terms of high-exponent f functions.

The effect is linked to the 3d, 4f, and 5f virtual orbitals of second, third, and fourth row elements approaching the valence shell as one approaches, respectively, the first-row transition metals, the lanthanides, and the actinides. Additionally, with increasing oxidation states, these orbitals will sink still lower and become still better back-donation acceptors from halogen and chalcogen lone pairs.

In the third row p-block, the 4f is still too remote, while the 4d is adequately covered by the basis functions needed to describe the 3d subvalence orbital.

An alternative explanation in terms of core polarization was refuted by means of large-core ECP calculations in which there are no core orbitals left to polarize.

## ■ ASSOCIATED CONTENT

### Supporting Information

The Supporting Information is available free of charge at https://pubs.acs.org/doi/10.1021/acs.jpca.3c00544.

Spreadsheet with XYZ coordinates and all calculated ΔTAE values and harmonic frequencies at the PW6B95 levels of theory, as well as the full ref 26 (XLSX)

## ■ AUTHOR INFORMATION

### Corresponding Authors

**Nisha Mehta** − *Department of Molecular Chemistry and Materials Science, Weizmann Institute of Science, 76100 Reḥovot, Israel;* orcid.org/0000-0001-7222-4108; Phone: +972-8-9342533; Email: nisha.mehta@weizmann.ac.il; Fax: +972-8-9343029

**Jan M. L. Martin** − *Department of Molecular Chemistry and Materials Science, Weizmann Institute of Science, 76100 Reḥovot, Israel;* orcid.org/0000-0002-0005-5074; Phone: +972-8-9342533; Email: gershom@weizmann.ac.il; Fax: +972-8-9343029

Complete contact information is available at: https://pubs.acs.org/10.1021/acs.jpca.3c00544

### Notes

The authors declare no competing financial interest.

## ■ ACKNOWLEDGMENTS

Work on this paper was supported by the Israel Science Foundation (grant 1969/20), by the Minerva Foundation (grant 2020/05), and by a research grant from the Artificial Intelligence and Smart Materials Research Fund (in memory of Dr. Uriel Arnon), Israel. Nisha Mehta would like to acknowledge the Feinberg Graduate School for a Sir Charles Clore Postdoctoral Fellowship as well as Dean of the Faculty and Weizmann Postdoctoral Excellence fellowships. The authors would like to thank Drs. Irena Efremenko and Mark A. Iron for assistance with the Table of Contents graphic and the NBO7 interface, respectively, and Mr. Emmanouil Semidalas for critical reading of the manuscript.

## ■ REFERENCES